\documentclass[twocolumn,showpacs,preprintnumbers,amsmath,amssymb]{revtex4}

\usepackage{graphicx}
\usepackage{dcolumn}
\usepackage{bm}
\usepackage{amsmath}
\usepackage{amssymb}

\begin{document}

\preprint{APS/123-QED}

\title{Simultaneous Excitation of Spins and Pseudospins in the Bilayer $\nu=1$ 
Quantum Hall State}

\author{D. Terasawa}
 \email{terasawa@lowtemp.phys.tohoku.ac.jp}
\author{M. Morino}%
\author{K. Nakada}
\author{S. Kozumi}
\author{A. Sawada}
\author{Z. F. Ezawa}
\affiliation{%
Department of physics, Tohoku University, Sendai 980-8578, Japan\\
}%

\author{N. Kumada}
\author{K. Muraki}
\author{Y. Hirayama}
\affiliation{
NTT Basic Research Laboratories, NTT corporation, Atsugi, Kanagawa 243-0198, Japan}%

\author{T. Saku}
\affiliation{
NTT-AT, Atsugi, Kanagawa 243-0198, Japan}%

\date{\today}

\begin{abstract}
The tilting angular dependence of the energy gap 
was measured in the bilayer 
quantum Hall state at the Landau level filling $\nu=1$ by changing
the density imbalance between the two layers.
The observed gap behavior shows a continuous transformation
from the bilayer balanced density state to the monolayer state.
Even a sample with 33 K tunneling gap shows the same activation energy anomaly
reported by Murphy {\it et al.} \cite{Murphy}.
We discuss a possible relation between our experimental results and
the quantum Hall ferromagnet of spins and pseudospins.
\end{abstract}

\maketitle

\section{Introduction}

In the $\nu=1$ bilayer quantum Hall (QH) state, 
when both layers have the same electron density
and sufficiently close together,
interlayer Coulomb interactions can produce
a many-body state
even in the absence of
interlayer tunneling \cite{Murphy,Boebinger,MacDonald}.
This collective state is thought to have a broken symmetry,
which is viewed as an easy-plane ferromagnet in the pseudospin space.
Pseudospin ``up'' (``down'') refers to an electron 
in the ``front'' (``back'') layer.
The huge tunneling conductance observed in this collective state
is considered as the Goldstone mode
of this broken symmetry \cite{Ezawa&Iwazaki,Spielman}.
Intriguingly,
this collective state has 
a remarkable tilting angular dependence
reported by Murphy {\it et al.} \cite{Murphy},
which is characterized as a rapid decrease of energy gap
before crossing over into a roughly angular independent region.
This dependence is explained 
as a commensurate-incommensurate (CIC) transition \cite{Yang},
a change in pseudospin ferromagnetic properties of the ground state.
The lowest-energy charged excitation 
has been considered
as a pair of pseudospin vortices 
called meron \cite{Moon}.
In contrast, the monolayer $\nu=1$ state is also
a broken symmetry ferromagnetic ground state.
In this case, the spontaneous ferromagnetic order is
in the spin space,
which leads to the system possessing an unusual
spin excitation known as skyrmion \cite{Sondhi,Fertig,Barrett,Schmeller}.
The energy gap increases with tilting
because the Zeeman energy, $g^\ast\mu_BB_{tot}$, 
is increased \cite{Schmeller}.
Here $g^\ast$ is the gyromagnetic ratio ($g^\ast = -0.44$),
$\mu_B$ is the Bohr magneton and $B_{tot}$ is the total
magnetic field.
However, 
except for those two extremes,
the bilayer $\nu=1$ state
has scarcely been investigated so far.
It is interesting 
how the energy gap changes 
as the density difference is made between the two layers.

In this paper, 
we report on the tilting angular dependence of
the activation energy in $\nu=1$ bilayer QH states,
especially when the electron densities
in the front layer $n_f$ and in the back layer $n_b$ are different.
Our earlier experiment has already revealed 
a consecutive existance of the $\nu=1$
bilayer QH state from the bilayer balanced density state 
to the monolayer limit \cite{Sawada},
but comprehensive experiments are yet to be performed.
Representing the normalized density difference as
$\sigma = ( n_f -n_b)/(n_f + n_b)$,
the energy gap shows a strong $\sigma$-dependent behavior
and the overall change exhibits a continuous transformation
from $\sigma=0$ to 1.
In addition,
we measured the energy gap of the $\nu=1$ bilayer $\sigma=0$ state
in a sample with  considerably large tunneling gap.

\section{Experimental result}
Samples used in this experiment
are GaAs/Al$_x$\\Ga$_{1-x}$As double-quantum-well
heterostructures grown by molcular beam epitaxy.
We used mainly two samples;
they both have two 20\,nm wide GaAs wells
separated by a 3.1\,nm Al$_x$Ga$_{1-x}$As barrier layer.
Their tunneling gap  $\Delta_{\rm SAS}$ 
are 1\,K ($x=1$) and 11\,K ($x=0.33$).
We additionally prepared a sample with extremely large
tunneling gap.
Having the barrier layer thickness of 1\, nm and
the Al concentration of 0.33,
the tunneling gap of this sample is 33\, K.
A unique feature of these sample structures is
that the modulation doping is carried out 
only on the front side of the double quantum well, 
and electrons in the other side of the layer 
is fully field-induced
by applying a positive bias to 
an underlying $n^+$ -GaAs
back gate.
This method enables us to control
the electron density
without deteriorating the mobility \cite{Muraki}.
The low temperature mobility
of all samples is $2 \times 10^6$\,cm$^2$/Vs,
except for $\Delta_{\rm SAS} = 33$\,K sample $5.4 \times 10^5$\,cm$^2$/Vs
at electron density $1.0 \times 10^{11}$ cm$^{-2}$.

Measurements were performed
with the sample mounted in a mixing chamber of
a dilution refrigerator.
Standard low-frequency ac lock-in techniques were used
with a current of 20\,nA to avoid heating effects.
A goniometer with
a superconducting stepper motor was used
to rotate the samples in the magnetic field \cite{Suzuki}.
The activation energy gap $\Delta$ was determined from
the temperature dependence of the magnetoresistance
$R_{xx} \sim \exp(-\Delta/2T)$.

\begin{figure}[t]
\begin{center}\leavevmode
\includegraphics[width=0.75\linewidth]{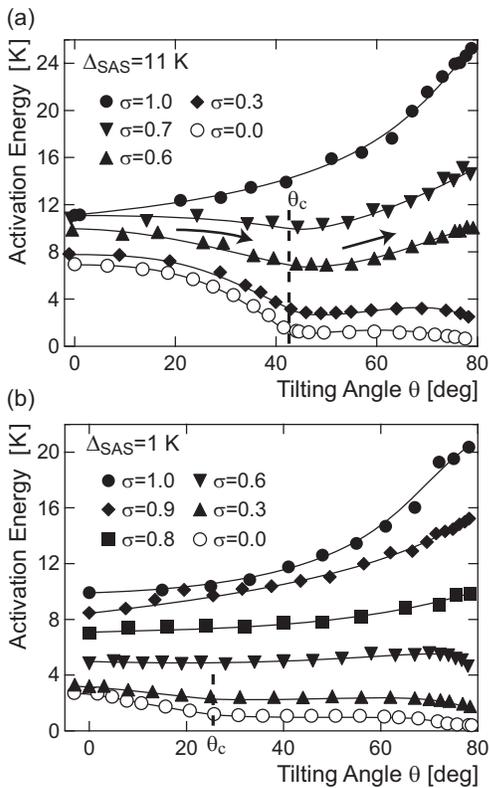}
\caption{\label{Graph_D-q} The angular dependence
of the activation energy
(a) for $\Delta_{\rm SAS} = 11$\,K and
(b) for $\Delta_{\rm SAS} = 1$\,K.
It looks like that the behavior at $\sigma=0$
transforms into that at $\sigma=1$
continuously in both samples.
The solid thin curves are guides to the eye.
}
\label{Graph_D-q}
\end{center}
\end{figure}

In Fig. \ref{Graph_D-q},
we show the angular dependence of the activation energy
at various density differences
(a) for $\Delta_{\rm SAS}=11$\,K and (b) for 1\,K.
Here $\theta$
is  the tilting angle of the magnetic field
from the normal line of the sample.
The total density ($n_f + n_b$) 
is fixed to 0.6 $\times 10^{11}$\, cm$^{-2}$.
In both samples,
the activation energy 
behavior at $\sigma=0$ is substantially the same as 
the result of Murphy {\it et al.} \cite{Murphy}.
The gaps drop until they reach the CIC transition angle $\theta_c$,
and then go into a roughly angle-independent regime.
The transition angle $\theta_c$ is 43$^\circ$ for 11\,K sample and 
27$^\circ$ for 1\,K sample.
At the monolayer point,
the gaps indicate a typical skyrmion-like behavior.
The number of flipped spins is approximately 7,
according to the method employed by Schmeller {\it et al.} \cite{Schmeller}.
Yet the most remarkable results of this measurement 
is that the energy gaps transform continuously 
from the balanced point to the monolayer limit.
It seems that there is no clear transition 
from meron-pair excitations to skyrmion excitations.
Furthermore,
the changes at $\sigma=0.6$ and 0.7 in the sample
of $\Delta_{\rm SAS}=11$\,K
are somewhat unexpected and yet more noteworthy.
They drop before the CIC transition angle,
which is not different from $\sigma=0$,
then they start increasing with further tilting.
The numbers of flipped spins in the region of starting to increase
are 2.4 at $\sigma=0.6$ and 2.5 at $\sigma=0.7$, 
indicating a possible excitation of skyrmions.

\begin{figure}[t]
\begin{center}\leavevmode
\includegraphics[width=0.8\linewidth]{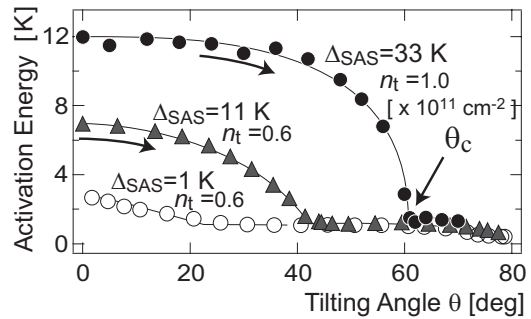}
\caption{Tilting angular dependence of the energy gap at $\sigma=0$,
for $\Delta_{\rm SAS} =$1, 11 and 33\,K sample.
The gap is diminished with tilting 
and the phase transition (indicated by the arrow) is observed.
The solid lines are guides to the eye.}
\label{D-q_s=0}
\end{center}
\end{figure}

The energy gap of the sample
with a large tunneling gap exhibited in Fig. \ref{D-q_s=0} is very intriguing.
We expected an increasing behavior
because the Zeeman gap $g\mu_BB_{tot}$ ($\approx 1$\,K) is much
smaller than the tunneling gap $\Delta_{\rm SAS}$ ($\approx 33$\,K),
which affects pseudospins as a `pseudomagnetic field' \cite{Muraki2}
in the one-particle state.
However, the observed angular dependence at the balance point
decreases by tilting
and shows the characteristic change of the meron-pair excitation.

\section{Discussions}

\begin{figure}[t]
\begin{center}\leavevmode
\includegraphics[width=0.9\linewidth]{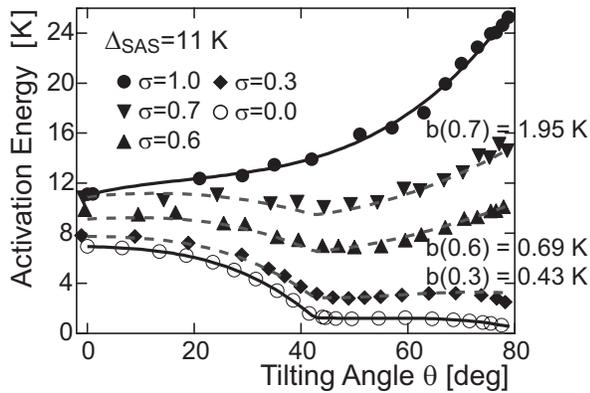}
\caption{The fitting result
of the tilting dependence of the activation energy.
The solid lines are obtained by 
polynomial fitting of 
$\Delta(0,\theta)$ and $\Delta(1,\theta)$ data. 
The broken lines indicate
$\Delta(\sigma,\theta)$ calculated by equation (\ref{fitting}).}
\label{Fit_D-q}
\end{center}
\end{figure}

Our experimental result is well
reproduced by the following equation [Fig.\ref{Fit_D-q}]
\begin{equation}
\Delta(\sigma,\theta)=(1-\sigma^2)\Delta(0,\theta)
+ \sigma^2\Delta(1,\theta) +b(\sigma),
\label{fitting}
\end{equation}
where $\Delta(\sigma,\theta)$ is
the activation energy at density difference $\sigma$
and tilting angle $\theta$.
Here $\Delta(0, \theta)$ and $\Delta(1, \theta)$
are  polynomial fit to the data at $\sigma=0$ and 1, respectively,
while $b(\sigma)$ is a phenomenological bias term.
Values of $b(\sigma)$ are described in the figure.
Here we show the result of the $\Delta_{\rm SAS} = 11$\,K sample only,
but we have obtained substantially the same fitting result 
for the $\Delta_{\rm SAS} = 1$\,K sample.

Equation (\ref{fitting}) indicates two essential points.
First, the excitation gap of the meron-pair
and that of the skyrmion 
are proportional to $1-\sigma^2$ and  $\sigma^2$,
respectively.
Second,
a quasiparticle must have the properties of both 
the meron-pair and the skyrmion
at an intermediate value of $\sigma$.

We argue the dependence of the excitation gap on $\sigma$
by studying how the spin and pseudospin stiffnesses depend on it.
By increasing $\sigma$ from zero,
the effects of the intralayer Coulomb interaction becomes larger
in the state dominated by the interlayer Coulomb exchange interactions.
Being accompanied by this Coulomb interactions,
the spin stiffness $\rho_s$, 
for being originated in the exchange energy between the spins,
increases as $\sigma^2$
because the exchange energy is 	 
proportional to the probability of 
the adjacent electron existance.
In contrast,
the meron-pair excitation gap becomes smaller
since the interlayer Coulomb interaction decreases
with increasing $\sigma$.
The pseudospin stiffness $\rho_{ps}$
is calculated within the Hartree-Fock Approximation as
$\rho_{ps} = (1-\sigma^2)\rho_E$,        
where $\rho_E$ is the interlayer exchange stiffness
when the layers are balanced\cite{Hanna}.
Therefore, the $\sigma$-dependence of the skyrmion excitation gap
and the meron-pair excitation gap
are expected to be proportional to $\sigma^2$ and $1-\sigma^2$, respectively.

\begin{figure}[b]
\begin{center}\leavevmode
\includegraphics[width=0.75\linewidth]{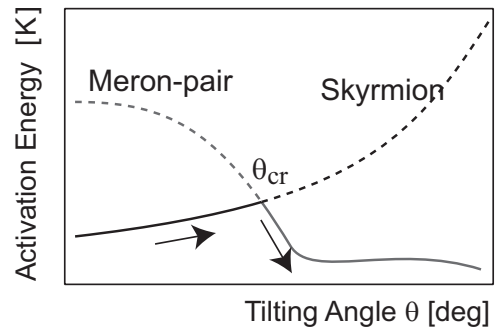}
\caption{An illustration of a would-be level crossing of a meron-pair
and a skyrmion excitations. The activation energy would realize
along the solid line. 
Actually, it looks like the experimantal results 
rather trace the broken line.}
\label{TypicalCrossing}
\end{center}
\end{figure}

We also have to consider the enhancement of 
the direct Coulomb energy
concomitant with the excitation.
This could be the origin of the term $b(\sigma)$ in equation (\ref{fitting}).
To suppress this Coulomb energy enhancement at the minimum,
both spins and pseudospins must be excited simultaneously.
Indeed, our experimental result shows that, in imbalanced
density states at $\nu=1$, the charged excitation
carries both the spin and pseudospin components.
The reason reads as follows.
Let us assume either the meron-pair or the skyrmion is excited.
If a level crossing occurs as the sample is tilted,
we can argue that it occurs only as in Fig.\ref{TypicalCrossing}.
In this case, a naive expectation is that
skyrmions and meron-pairs are excited for $\theta<\theta_{cr}$ 
and $\theta>\theta_{cr}$ with a certain critical value $\theta_{cr}$, 
respectively. 
Then the activation energy must be realized
along the solid line in the figure. 
On the contrary, our experimental result is against a simple
level crossing between a skyrmion and a meron-pair excitations.
Thus, to realize the experimental curve,
it is necessary that there exists a single excitation carrying
both spins and pseudospins which is reduced to the skyrmion 
in the monolayer limit ($\sigma=1$) and to the meron-pair 
at the balanced point ($\sigma=0$).
Such a simultaneous flip of spins 
and pseudospins may be the predicted SU(4) skyrmion
in the $\nu=1$ bilayer QH state \cite{Ezawa}.

Finally, the present experiment suggests that
pseudospins are excited at $\sigma=0$ even in a sample with
very large tunneling gap [Fig. \ref{D-q_s=0}].
An intriguing behavior of this activation energy
as a function of the tilting angle in Fig. \ref{D-q_s=0}
is yet to be explained.

In conclusion,
we measured the tilting angular
dependence of the energy gap by changing the density difference
in the $\nu=1$ bilayer QH state.
We have found a simultaneous excitation of spins and pseudospins
in imbalanced density states.

\begin{acknowledgments}
This research was supported in part by
Grants-in-Aid for the Scientific Research from 
the Ministry of Education,
Culture, Sports, Science and Technology of Japan (Nos. 14010839,14340088) and
CREST-JST.
\end{acknowledgments}

\end{document}